\documentclass[aps,prl,preprint,grouppedaddress,showpacs]{revtex4}

\usepackage{graphicx}%
\usepackage{dcolumn}
\usepackage{amsmath}
\usepackage{multirow}
\usepackage{epsfig}
\usepackage{amssymb}
\usepackage{pifont}
\usepackage{color}

\begin{document}

\title{Rich variety of defects in ZnO via an attractive interaction
between O-vacancies and Zn-interstitials}

\author{Yong-Sung Kim}
\email{yongsung.kim@kriss.re.kr}
\affiliation{Korea Research Institute of Standards and Science, P.O. Box 102, Yuseong, Daejeon 305-600, Korea}
\author{C. H. Park}
\email{cpark@pusan.ac.kr}
\affiliation{Research Center for Dielectric and Advanced Matter Physics, Pusan National University, Pusan 609-735, Korea}

\date{\today}% It is always \today, today, but you may specify any date with \date.

\begin{abstract}
As the concentration of intrinsic defects becomes sufficiently high in O-deficient ZnO, interactions between defects lead to a significant reduction in their formation energies.
We show that the formation of both O-vacancies and Zn-interstitials becomes significantly enhanced by a strong {\it attractive} interaction between them, making these defects an important source of $n$-type conductivity in ZnO.
\end{abstract}

%PACS, the Physics and Astronomy Classification Scheme.
\pacs{71.55.Gs, 72.80.Ey, 73.61.Ga}

\maketitle

It is well-known that O-deficient ZnO can easily become $n$-type even without the introduction of any intentional dopants. The mechanisms leading to the $n$-type behavior are, however, still controversial.
Even though native defects resulting from thus deficiency have been excluded as the main source of the high free electron density, the $n$-type conductivity of ZnO is seen to be closely related to its O-deficiency which manifests itself through the formation of O-vacancy (V$_{\rm O}$) and/or Zn-interstitial (I$_{\rm Zn}$) defects.

Theoretical first-principles studies of the formation enthalpies of point defects in ZnO have indicated that neither V$_{\rm O}$ nor I$_{\rm Zn}$ can lead to a high concentration of free carriers, since the most stable donor-like defect V$_{\rm O}$ has been shown to be a deep donor and the formation of the I$_{\rm Zn}$ shallow donor state has been known to be energetically far less favorable than that of V$_{\rm O}$ when the Fermi level is close to the conduction band minimum (CBM) \cite{VW00b, CH01, ZU01, OB01, VW05, AL05, VW07}.
On the other hand, hydrogen contamination was proposed to be an important cause of the natural $n$-doping \cite{VW00l,VW03,VW07N}.
A potential problem with this proposal is that a high concentration of electron carriers is still observed even when H contamination is minimized or when H is removed \cite{Halliburton,OG00,CH03}.
The sample annealed at 1100 $^\circ$C in Zn vapor had an electron concentration of 1.5$\times$10$^{18}$ cm$^{-3}$,
while the untreated sample had 1.3$\times$10$^{17}$ cm$^{-3}$ \cite{Halliburton}.
Hydrogen easily diffuses out of ZnO at high temperature, and
the H-related hyperfine structure
observed in H-contaminated samples from the electron-nuclear-double-resonance study \cite{Hofmann}
was not observed in such heat-treated samples \cite{HalliburtonPrivate}.
A meta-stable shallow donor state of V$_{\rm O}$ was suggested as an alternative source of the $n$-doping \cite{ZU05}, but it is still controversial.

As the concentration of defects becomes high, interactions between defects become unavoidable and they need to be taken into account. It is known that donor-like and acceptor-like defects are attracted towards each other through their Coulombic interaction \cite{LI98}.
However, the defect-defect interactions between donor-like defects have been neglected in previous theoretical studies of ZnO.
In this Letter, we show that there can be a strong {\it attractive} interaction between two donor-like defects, the deep-donor V$_{\rm O}$ and the shallow-donor I$_{\rm Zn}$. The driving force for the attractive interaction is the quantum mechanical hybridization between the electronic orbitals of their respective deep and shallow donor states, which attracts I$_{\rm Zn}$ toward V$_{\rm O}$. The interaction significantly lowers the energy of the electronic donor orbital of V$_{\rm O}$, as the distance between two defects decreases. The overall effect is a large reduction in the total energy of the system. As a result, the concentration of I$_{\rm Zn}$ can reach a high enough level to explain the high concentration of electron carriers in O-deficient ZnO, even if the Fermi level is close to the CBM.

%We performed density functional theory calculations with projector augmented wave (PAW) pseudopotentials \cite{BL94}, as implemented in the Vienna {\it Ab Initio} Simulation Package (VASP) code \cite{KR99}. A plane-wave cutoff of 495 eV, a 2$\times$2$\times$2 special ${\bf k}$-points mesh in the Brillouin zone, and a supercell containing 96 atoms in wurtzite structure are used. The local-density-functional approximation \cite{CA80} with on-site Coulomb interaction correction \cite{AN93} for the Zn 3$d$ states (LDA+$U$) is used for the exchange-correlation energy. A 6 eV value for $U$ was chosen. The theoretical equilibrium lattice constants of $a$=3.127 \AA~with $c/a$=1.602 are used in the overall calculations.
%The calculated LDA+$U$ gap is 1.74 eV, and the V$_{\rm O}$ level is estimated to be at 0.4 eV below the CBM, which is consistent with an experimental data of thermal excitation of carrier []. Since the orbital characters of V$_{\rm O}$ and I$_{\rm Zn}$ are Zn-4$s$-like similarly to the CBM character, the level position determined from the CBM is considered to be more reasonable. Here, we focused on the interactions between defects, independent of the absolute values of the formation enthapies of the defects themselves.

\begin{figure}[b]
\includegraphics[width=0.55\linewidth]{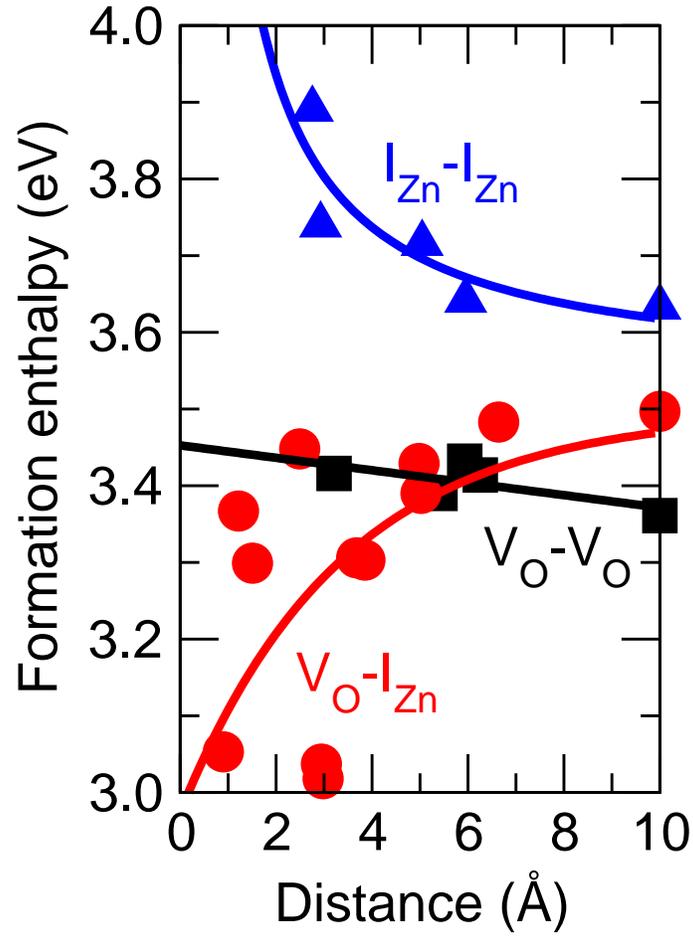}
\caption{(color online). Calculated formation enthalpies of the defect pairs as a function of the distance.
Attractive interaction of the V$_{\rm O}$-I$_{\rm Zn}$ pair (\ding{108}), and repulsive interactions between the same types of defects, V$_{\rm O}$-V$_{\rm O}$ (\ding{110}) and I$_{\rm Zn}$-I$_{\rm Zn}$ (\ding{115}) are shown.} \label{fig1}
\end{figure}

We performed density functional theory calculations, as implemented in the Vienna {\it Ab Initio} Simulation Package (VASP) code \cite{KR99,aux}.
In order to examine interactions between defects, we employ a supercell geometry, in which two defects are located at various sites in the cell, and calculate the variation of the total-energy depending on their distance.
We consider two low-energy donor-like defects: V$_{\rm O}$ and I$_{\rm Zn}$.
In $n$-type O-deficient ZnO, the Fermi level is close to the CBM, and the neutral charge state of V$_{\rm O}$ is stable, since V$_{\rm O}$ is a deep-donor defect, but the shallow-donor I$_{\rm Zn}$ is stable in a (2+)-charge state.
Therefore, only interactions between these charge states are focused on in this work. The results of our calculations are shown in Fig. \ref{fig1}, where the variations of formation enthalpies [$\Omega_{\alpha,\alpha'}(r)$] as a function of defect-defect ($\alpha$-$\alpha'$) separation ($r$) are shown. The Fermi level is assumed to be located at the V$_{\rm O}$ deep donor level.

We first consider interactions between two V$_{\rm O}$'s and, separately, between two I$_{\rm Zn}$'s. It is found that the total energy increases as the distance between the defects decreases, indicating a repulsive interaction between same-type defects. As shown in Fig. \ref{fig1}, the energy of two I$_{\rm Zn}$'s increases sharply as they get closer to each other, whereas that of two V$_{\rm O}$'s increases slightly. The repulsive interaction between I$_{\rm Zn}$'s is due to the expected Coulombic repulsion. The variation of the energy is fit well using the screened Coulombic interaction functional, $\frac{Z^2e^2}{4\pi\epsilon}\cdot\frac{e^{-r/\lambda}}{r}$, where the fitting parameters are the charge state $Z$=2, the dielectric constant $\epsilon$=8.65, and the screening length $\lambda$=1.13 \AA. The weak repulsive interaction between V$_{\rm O}$'s is related to a strain effect. The inward-relaxation of neutral V$_{\rm O}$ induces a tensile-strain into the surrounding lattice. In the paired state, the strain does not become fully relaxed.

\begin{figure}[bp]
\includegraphics[width=0.75\linewidth]{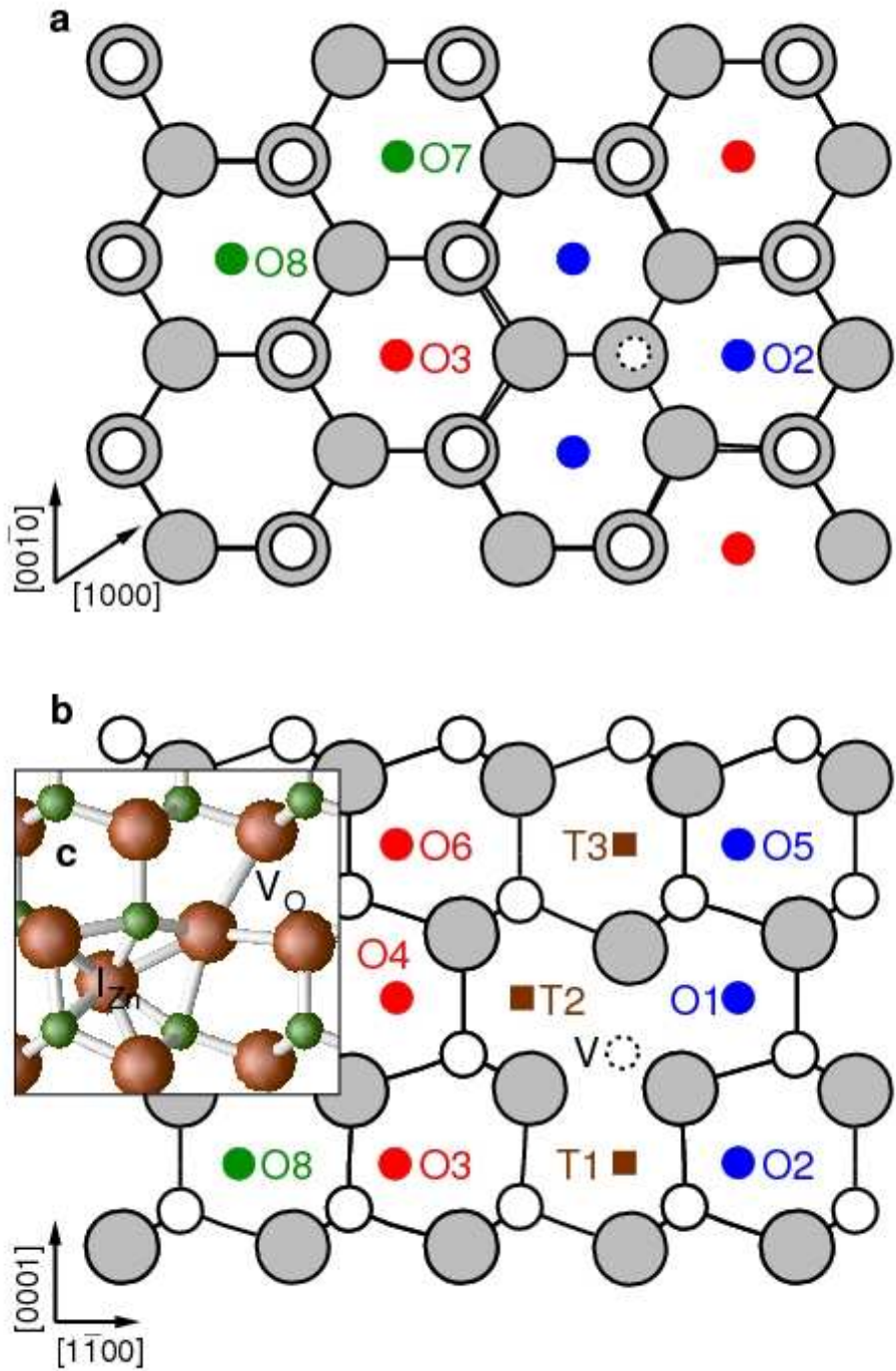}\\
\caption{(color online). Available atomic sites for an I$_{\rm Zn}$ near a V$_{\rm O}$ are indicated
in the ZnO wurtzite structure shown from (a) the top (along the polar axis) and (b) from the side view.
(c) Atomic structure of the most stable V$_{\rm O}$-I$_{\rm Zn}$ pair ({\sf O3}). The large and small atoms indicate the Zn and O atoms, respectively.} \label{fig2}
\end{figure}

An interesting finding is that there is a strong attractive interaction between V$_{\rm O}$ and I$_{\rm Zn}$. It is surprising that an attractive interaction is induced between two donor-like defects. The formation enthalpy of V$_{\rm O}$-I$_{\rm Zn}$ pairs as a function of their separation is shown in Fig. \ref{fig1}. The total-energy decreases significantly, as the separation becomes smaller, indicating that I$_{\rm Zn}$ is stable with respect to a coupling to an on-site Zn atom surrounding V$_{\rm O}$. In Fig. \ref{fig2}(a) and (b), the possible sites for I$_{\rm Zn}$ around V$_{\rm O}$ are indicated; the octahedral interstitial sites (${\sf O1}$-${\sf O8}$), the tetrahedral interstitial sites ({\sf T1}-{\sf T3}), and the V$_{\rm O}$ site ({\sf V}). The most stable site is the {\sf O3} octahedral interstitial site for which the atomic structure is shown in Fig. \ref{fig2}(c). The binding energy is 0.52 eV. There are three equivalent {\sf O3} sites around a V$_{\rm O}$. When I$_{\rm Zn}$ is trapped just at the oxygen vacancy site {\sf V} (as an antisite defect), the 0.44 eV binding energy is slightly smaller.

\begin{figure}[b]
\includegraphics[width=1.00\linewidth]{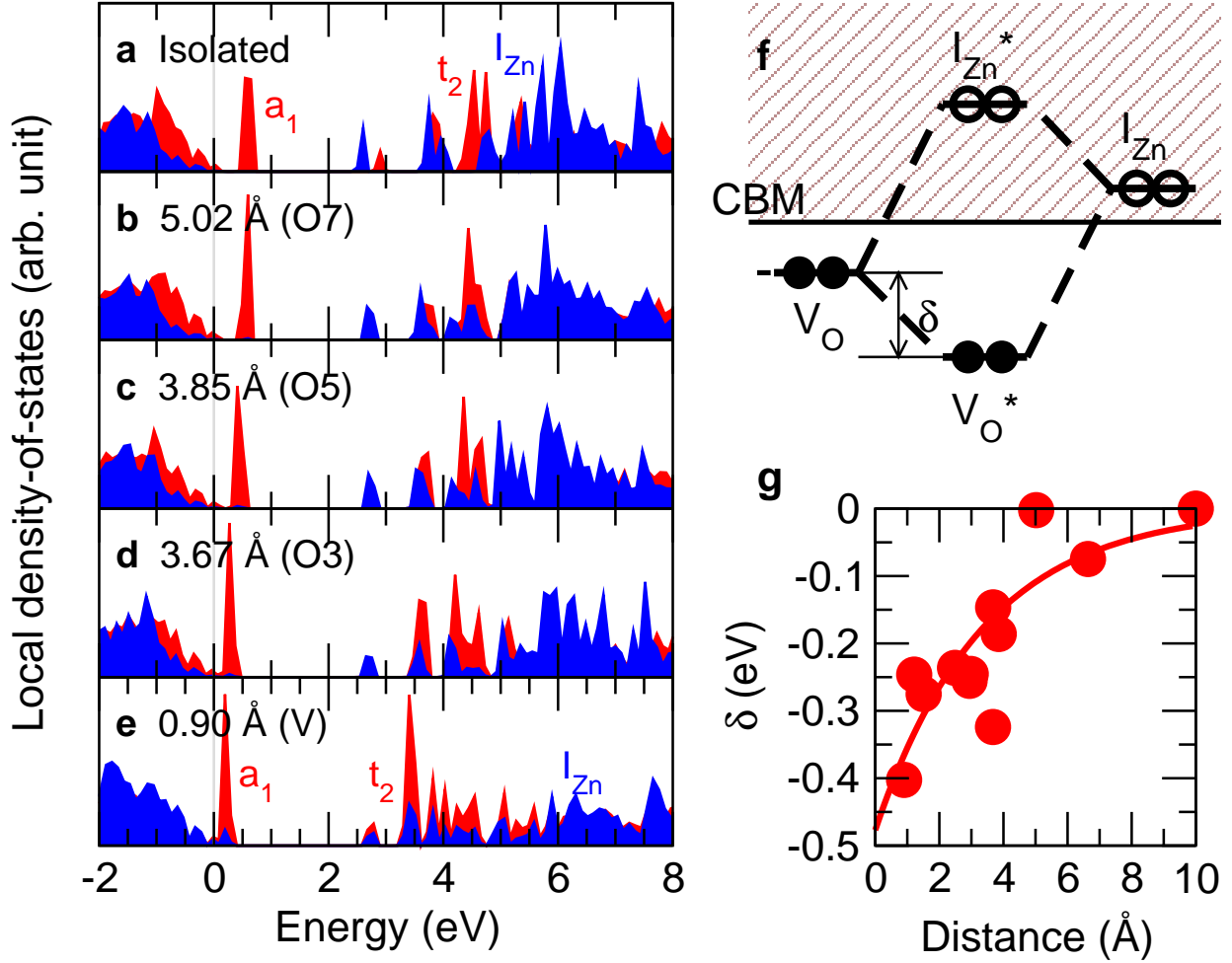}\\
\caption{(color online). Calculated local density-of-states of electrons near V$_{\rm O}$ (red) and near I$_{\rm Zn}$ (blue) for (a) the separated V$_{\rm O}$ and I$_{\rm Zn}$ (a), and for the V$_{\rm O}$-I$_{\rm Zn}$ pairs of (b) {\sf O7}, (c) {\sf O5}, (d) {\sf O3}, and (e) {\sf V} of which sites are indicated in Fig. \ref{fig2}(a) and (b).
(f) Schematic diagram of the defect levels of V$_{\rm O}$, I$_{\rm Zn}$, and V$_{\rm O}$-I$_{\rm Zn}$. V$_{\rm O}^*$ and I$_{\rm Zn}^*$ denote the V$_{\rm O}$ and I$_{\rm Zn}$ in V$_{\rm O}$-I$_{\rm Zn}$ pair, respectively.
(g) Calculated level drops $\delta$ of the V$_{\rm O}$ state as a function of the distance.} \label{fig3}
\end{figure}

In order to understand the driving force of the attraction between V$_{\rm O}$ and I$_{\rm Zn}$, we examined the change of the electronic structure, depending on distance. The electronic structure of the coupled state is significantly modified as the separation between the two defects becomes smaller. It is noted that the defect levels of the V$_{\rm O}$ (both the occupied $a_1$ gap state and the unoccupied $t_2$ state) are lowered while those of I$_{\rm Zn}$ state are raised. The calculated results for the electronic structure are described by the local density-of-states from the four Zn atoms around V$_{\rm O}$ and from the I$_{\rm Zn}$ atom, as shown in Fig. \ref{fig3}(a)-(e). Figure \ref{fig3}(g) describes the lowering of the V$_{\rm O}$ state as a function of the distance. The lowering is noted to be quantitatively similar to the change of the formation enthalpy in Fig. \ref{fig1}. This indicates that there is a strong interaction between these two defect-levels.

The driving force for the attraction originates from the hybridization between the deep donor state and the shallow donor state. The change in electronic structure can be simply depicted as in Fig. \ref{fig3}(f). The defect level of the shallow donor I$_{\rm Zn}$ is located inside the conduction bands and the V$_{\rm O}$ level is located deep within the band gap. Both defect orbitals mainly come from the Zn-4$s$ orbitals. As the distance between the two defects becomes smaller, a hybridization between the two defect orbitals is induced, as a result of which the lower-lying V$_{\rm O}$ level is lowered, and the upper I$_{\rm Zn}$ level is pushed-up. Since the V$_{\rm O}$ level is occupied by two electrons while the I$_{\rm Zn}$ level is empty, the hybridization lowers the total energy. As the defect-defect separation becomes smaller, the hybridization becomes stronger. This gives rise to an attractive interaction between the deep and shallow donor states. As shown in Fig. \ref{fig3}(g), the change of the V$_{\rm O}$ level is well-fit by an exponential function: $\delta_0 e^{-r/a^*}$, with the fitting parameters, $\delta_0$=0.47 eV, which is $\delta$ at $r$=0 \AA, and $a^*$=4.29 \AA. The wave functions of the defect levels of the donor-like V$_{\rm O}$ and I$_{\rm Zn}$ can be approximately written by $e^{-r/a_{\rm V}}$ and $e^{-r/a_{\rm I}}$, respectively, where $a_{\rm V}$ and $a_{\rm I}$ are the effective Bohr radii of the defect electron densities. The deep level of V$_{\rm O}$ is localized nearly within the second oxygen nearest neighbor, $a_{\rm V}$$\sim$5.5 \AA, and the shallow donor state is delocalized as $a_{\rm I}$=$a_0\epsilon$/$m_e^*$=20 \AA~ within the simple effective mass theory, where the Bohr radius $a_0$=0.53 \AA, the dielectric constant $\epsilon$=8.65, and the effective mass of electron $m_e^*$=0.23. The fitting value of $a^*$ can be obtained from the product of the two wave functions,  $\langle\psi_{\rm V}|\psi_{\rm I}\rangle$, with $a_{\rm V}$=5.5 \AA~and $a_{\rm I}$=20 \AA, i.e., 1/$a^*$=1/$a_{\rm V}$+1/$a_{\rm I}$=1/(4.3 \AA). There is just a slight deviation between the lowering of the total energy (Fig. \ref{fig1}) and the V$_{\rm O}$ level [Fig. \ref{fig3}(g)] from the fitting curves, which is related to the local strain effect. The closest V$_{\rm O}$-I$_{\rm Zn}$ pair ({\sf V}) shows the largest level drop 0.403 eV [see Fig. \ref{fig3}(e) and (g)], and the lowest energy configuration is found at the {\sf O3} site, because the {\sf O3} interstitial volume is more spacious as it is located behind the inward-relaxed V$_{\rm O}$ [see Fig. \ref{fig2}(c)], and thus the accommodation of I$_{\rm Zn}$ is easier. The interaction via orbital hybridization between the defects is more significant than the elastic interaction.

The attractive interaction between V$_{\rm O}$ and I$_{\rm Zn}$ can lead to the coexistence state of both V$_{\rm O}$ and I$_{\rm Zn}$ rather than the presence of only V$_{\rm O}$ in O-deficient ZnO. We estimated the concentrations of V$_{\rm O}$-I$_{\rm Zn}$ pair, relatively to isolated defects, and also the Fermi level through the calculations of the formation enthalpies of defects under the thermal equilibrium condition. The average formation enthalpy of a defect $\alpha$ under the interaction with $\alpha'$-defects is calculated by
\begin{equation}
\Omega_{\alpha}=\Omega_{\alpha}^0+\sum_{r}\sum_{\alpha'}U_{\alpha,\alpha'}(r),
\label{eqn:eform}
\end{equation}
where $\Omega_{\alpha}^0$ is the formation enthalpy in the isolated state without interaction, and $U_{\alpha,\alpha'}(r)$ is the interaction energy between $\alpha$- and $\alpha'$-defects separated by $r$. In Fig. \ref{fig1}, the calculated $\Omega_{\alpha,\alpha'}(r)=\Omega_{\alpha}^0+\Omega_{\alpha'}^0+U_{\alpha,\alpha'}(r)$ are shown, by which $U_{\alpha,\alpha'}(r)$ can be estimated. For the numerical simplicity, particularly in the limit of extreme high concentrations of defects, we adopt a mean-field concept, and $\Omega_{\alpha}$ can be approximated by $\Omega_{\alpha}^0+\sum_{\alpha'}\bar U_{\alpha,\alpha'}$, where
$\bar U_{\alpha,\alpha'}=\int dV U_{\alpha,\alpha'}(r)n_{\alpha'}$
is the mean interaction energy between defect $\alpha$ and $\alpha'$, and $U_{\alpha,\alpha'}(r)$ is approximated by the fitting curves drawn in Fig. \ref{fig1}. Here $n_{\alpha'}$ is the density of a defect $\alpha'$. When we employed the calculated values for $U_{\alpha,\alpha'}(r)$ shown in Fig. \ref{fig1}, the overall results show little change. Now we follow conventional method \cite{ZH91} to determine the concentration of each defects by
$n_{\alpha}=N_0 e^{-\beta\Omega_{\alpha}}$, the Fermi level ($\varepsilon_{\rm F}$), and the carrier concentration ($n_e$) by charge neutrality condition. $N_0$ is the number of available sites for defect formation: about 4.22$\times$10$^{22}$ cm$^{-3}$ in ZnO.
The concentration of electron carrier depending on $\varepsilon_{\rm F}$ is determined by the effective density of electronic states of the CBM ($N_c$) of 2.8$\times$10$^{18}$ cm$^{-3}$ at room temperature and a conduction-band electron effective mass of $m_e^*$=0.23. Since the concentration of hole carriers is minor, so the results are insensitive to the density of the state around the valance band maximum, we used an effective mass of $m_h^*$=0.5 \cite{LA02}. The formation enthalpies of the isolated V$_{\rm O}^0$ and I$_{\rm Zn}^{2+}$ are 1.68 eV and 2$\varepsilon_{\rm F}$-0.91 eV, respectively, under the Zn-rich condition, where the Zn chemical potential is chosen to be the energy of a Zn atom in the Zn hcp metal \cite{aux}. The crystal growth condition varied between O-rich and Zn-rich conditions is represented by the O chemical potential, which affects the defect concentration.

\begin{figure}[t]
\includegraphics[width=1.00\linewidth]{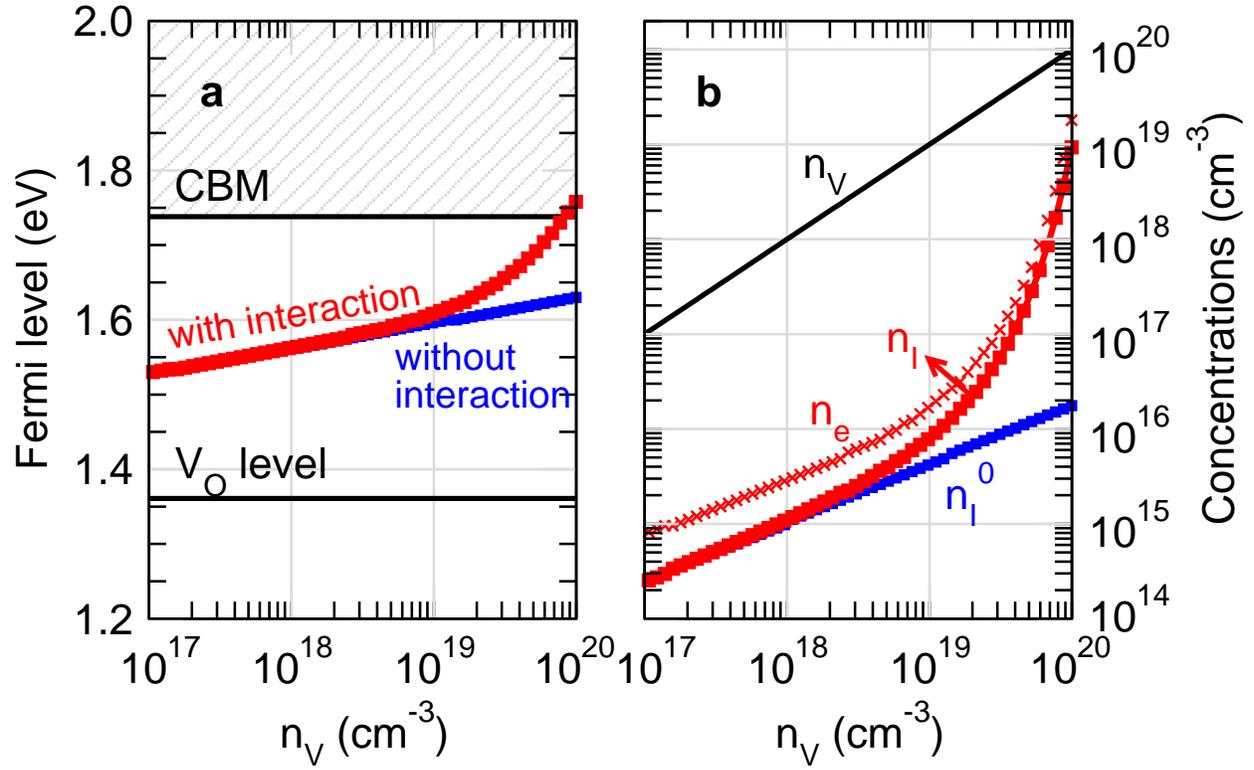}\\
\caption{(color online). (a) Fermi level and (b) equilibrium concentrations of I$_{\rm Zn}$ calculated by
considering the defect interactions (red) and neglecting the interactions (blue) as a function of V$_{\rm O}$ concentrations.
Theoretical values of the CBM and the V$_{\rm O}$ level are indicated in (a).
Equilibrium electron carrier concentrations ($n_e$) are also plotted in (b).}
\label{fig4}
\end{figure}

In Fig. \ref{fig4}, the concentration of I$_{\rm Zn}$ ($n_{\rm I}$) and the Fermi level are shown with respect to the concentration of V$_{\rm O}$ ($n_{\rm V}$) at a growth temperature of 900 K. When we neglect the interactions between defects, the Fermi level is calculated to be pinned at a mid-level between the CBM and the V$_{\rm O}$ level, and $n_{\rm I}$ is limited to much less than 10$^{17}$ cm$^{-3}$, as  conventionally indicated. When we include the interactions between defects, the result is significantly changed. Under O-deficient condition, $n_{\rm I}$ can become close to $n_{\rm V}$, i.e., the coexistence of the V$_{\rm O}$ and I$_{\rm Zn}$ in the O-deficient ZnO becomes remarkable. Since I$_{\rm Zn}$ is a shallow-donor defect, Fermi level can be raised to levels much higher than those expected in the presence of only V$_{\rm O}$'s and thus the O-deficient ZnO can be heavily $n$-type doped. These findings show how {\it the O-deficiency itself can be the source of the $n$-type conductivity in ZnO}.
%It is noted that, as $n_{\rm V}$ becomes higher than 10$^{19}$ cm$^{-3}$ and as $n_{\rm I}$ increases sharply with  $n_{\rm V}$, the effect of the repulsive interaction between I$_{\rm Zn}$'s is also emerges. However, its effect is far weaker than that of the interaction between V$_{\rm O}$ and I$_{\rm Zn}$, and the effect of the interaction between V$_{\rm O}$'s is also relatively minor.
When $n_{\rm V}$ rises up approximately 10$^{20}$ cm$^{-3}$, where the average distance between V$_{\rm O}$'s is about 27 \AA, $n_{\rm I}$ can reach about 10$^{19}$ cm$^{-3}$, and the Fermi level approaches the CBM with an electron carrier concentration $n_e$ of 2$\times$10$^{19}$ cm$^{-3}$.

Recently, Vlasenko and Watkins have reported the existence of a complex defect consisting of a deep donor V$_{\rm O}$ and a shallow donor in ZnO through the optical detection of electron paramagnetic resonance (EPR) in electron-irradiated ZnO \cite{VL05}. They identified an EPR line whose $g$ value is accurately given by the average of the $g$ values of the EPR signals from V$_{\rm O}$ and a shallow donor. The coincidence was suggested to be the result expected for a closely located defect pair, i.e., V$_{\rm O}$ and a shallow-donor. The pair may be explained by the V$_{\rm O}$-I$_{\rm Zn}$ pair reported in this present study.
It should be also noted that the defect level of V$_{\rm O}$ can be deeper below the midgap of ZnO when coupled with I$_{\rm Zn}$ than isolated.

In conclusion, we find a strong attractive interaction between V$_{\rm O}$ and I$_{\rm Zn}$, whose driving force originates mainly from the hybridization between the orbitals of the V$_{\rm O}$ deep and the I$_{\rm Zn}$ shallow donor states.
An I$_{\rm Zn}$ defect preferentially forms around a V$_{\rm O}$ defect as a result of the attractive interaction.
We find that in seriously O-deficient ZnO, the concentrations of V$_{\rm O}$ and I$_{\rm Zn}$ can be comparable.
Therefore, O-deficiency itself can be the source of the $n$-type conductivity in ZnO.

\begin{acknowledgments}
Y.S.K. acknowledges the support of KRCF through the `Development of Advanced Materials Metrology' project and also the support from the Korea Institute of Science and Technology Information under `The 11th Strategic Supercomputing Support Program'. C.H.P. acknowledges the support of the KRF through the `Important Research Center'. We also acknowledge Dr. Chadi for valuable comments.
\end{acknowledgments}

\newpage

%\begin{center}
%\Large Figure Captions
%\end{center}

\begin{thebibliography}{99}
\bibitem{VW00b}	A. F. Kohan, G. Ceder, D. Morgan, and C. G. Van de Walle,
                \prb {\bf 61}, 15019 (2000).
\bibitem{CH01}  E.-C. Lee, Y.-S. Kim, Y.-G. Jin, and K. J. Chang,
                \prb {\bf 64}, 085120 (2001).
\bibitem{ZU01}	S. B. Zhang, S.-H. Wei, and A. Zunger,
                \prb {\bf 63}, 075205 (2001).
\bibitem{OB01}	F. Oba, S. R. Nishitani, S. Isotani, H. Adachi, and I. Tanaka,
                \apl {\bf 90}, 824 (2001).
\bibitem{VW05}	A. Janotti and C. G. Van de Walle,
                \apl {\bf 87}, 122102 (2005).
\bibitem{AL05}	P. Erhart, A. Klein, and K. Albe,
                \prb {\bf 72}, 085213 (2005).
\bibitem{VW07}  A. Janotti and C. G. Van de Walle,
                \prb {\bf 76}, 165202 (2007).
\bibitem{VW00l}	C. G. Van de Walle,
                \prl {\bf 85}, 1012 (2000).
\bibitem{VW03}	C. G. Van de Walle and J. Neugebauer,
                Nature {\bf 423}, 626 (2003).
\bibitem{VW07N} A. Janotti and C. G. Van de Walle,
                Nature Mat. {\bf 6}, 44 (2007).
\bibitem{Halliburton}  L. E. Halliburton, N. C. Giles, N. Y. Garces, M. Luo,
                C. Xu, L. Bai, and L. A. Boatner,
                \apl {\bf 87}, 172108 (2005).
\bibitem{OG00}  K. Ogata, K. Sakurai, Sz. Fujita, Sg. Fujita, and K. Matsushige,
                J. Cryst. Growth {\bf 214-215}, 312 (2000).
\bibitem{CH03}  Z. Q. Chen, S. Yamamoto, M. Maekawa, A. Kawasuso, X. L. Yuan,
                and T. Sekiguchi,
                J. Appl. Phys. {\bf 94}, 4807 (2003).
\bibitem{Hofmann}  D. M. Hofmann, A. Hofstaetter, F. Leiter, H. Zhou, F. Henecker,
                B. K. Meyer, S. B. Orlinskii, J. Schmidt, and P. G. Baranov,
                \prl {\bf 88}, 045504 (2002).
\bibitem{HalliburtonPrivate} L. E. Halliburton, private communication.
\bibitem{ZU05}	S. Lany and A. Zunger,
                \prb {\bf 72}, 035215 (2005).
\bibitem{LI98}	S. Limpijumnong, S. B. Zhang, S.-H. Wei, and C. H. Park,
                \prl {\bf 92}, 155504 (2004).
%\bibitem{BL94}  P. E. Bl${\rm \ddot o}$chl,
%                \prb {\bf 50}, 17953 (1994).
\bibitem{KR99}  G. Kresse and D. Joubert,
                \prb {\bf 59}, 1758 (1999).
\bibitem{aux}  See EPAPS Document No. [] for more detail about the calculation method and
               some results for the isolated defects.
               For more information on EPAPS, see http://www.aip.org/pubservs/epaps.html.
%\bibitem{CA80}  D. M. Ceperley and B. J. Alder,
%                \prl {\bf 45}, 566 (1980).
%\bibitem{AN93}  V. I. Anisimov, I. V. Solovyev, M. A. Korotin,
%                M. T. Czy${\rm \dot z}$yk, and G. A. Sawatzky,
%                \prb {\bf 48}, 16929 (1993).
\bibitem{ZH91}	S. B. Zhang and J. E. Northrup,
                \prl {\bf 67}, 2339 (1991).
%\bibitem{SH06}	S. Shokhovets, G. Gobsch, and O. Ambacher,
%                Superlattices and Microstructures {\bf 39}, 299 (2006);
%                $m_h^*$ reported are scarce in the range of 0.31-0.59.
\bibitem{LA02}	W. R. L. Lambrecht, A. V. Rodina, S. Limpijumnong,
                B. Segall, and B. K. Meyer,
                \prb {\bf 65}, 075207 (2002).
\bibitem{VL05}	L. S. Vlasenko and G. D. Watkins,
                \prb {\bf 71}, 125210 (2005).
\end{thebibliography}
\end{document}